\documentclass[aps,showpacs]{revtex4}

\usepackage{graphicx}

\begin{document}


\newcommand{\be}{\begin{equation}}
\newcommand{\ee}{\end{equation}}
\newcommand{\pu}{\Psi_{\uparrow}}
\newcommand{\pd}{\Psi_{\downarrow}}
\newcommand{\p}{\phi}
\newcommand{\hb}{\hbar}
\newcommand{\R}{\Re}
\newcommand{\I}{\Im}
\newcommand{\A}{\theta}
\newcommand{\U} {\uparrow}
\newcommand{\D}{\downarrow}
\newcommand{\lar}{\longrightarrow}
\newcommand{\ket}[1]{\left| #1 \right>} 
\newcommand{\bra}[1]{\left< #1 \right|} 
\newcommand{\braket}[2]{\left< #1 \vphantom{#2} \right|
 \left. #2 \vphantom{#1} \right>} 
 \newcommand{\parenthnewln}{\right.\\&\left.\quad\quad{}}

\title{Quantum measurement and the Aharonov-Bohm effect with superposed magnetic fluxes}


\author{Ka\'{c}a Bradonji\'{c}}
\affiliation{Physics Department, Wellesley College, 106 Central Street, Wellesley, MA 02481}
\email[]{kbradonjic@gmail.com}

\author{John D. Swain}
\affiliation{Physics Department, Northeastern University, 110 Forsyth St., 111 Dana Research Center Boston, MA 02115}
\email[]{John.Swain@cern.ch}




\begin{abstract}
We consider the magnetic flux in a quantum mechanical superposition of two values and find that the Aharonov-Bohm effect interference pattern contains information about the nature of the superposition, allowing information about the state of the flux to be extracted without disturbance. The information is obtained without transfer of energy or momentum and by accumulated nonlocal interactions of the vector potential $\vec{A}$ with many charged particles forming the interference pattern, rather than with a single particle. We suggest an experimental test using already experimentally realized superposed currents in a superconducting ring and discuss broader implications.
\keywords{quantum measurement \and Aharonov-Bohm  \and magnetic flux}

\end{abstract}

\pacs{73.23.-b, 03.65.Ta} 


\maketitle

\section{Introduction\label{Introduction}}

The Aharonov-Bohm (A-B) effect involves the quantum mechanical scattering of electrons in the presence of a classical magnetic vector potential produced by a current-carrying solenoid. Electrons are prevented from entering the region where the magnetic field itself is nonzero so there is no classical force on them. Nevertheless, information about the flux can be obtained. Here, motivated by the experimental realization of macroscopic superpositions of classical currents in opposite directions around a superconducting ring \cite{vanderWal2000,Friedman}, we consider what happens to the A-B effect if the flux is in superposition of two states with different classical values. In distinction to references \cite{Anandan,Tsomokos1,Tsomokos2,Tsomokos3,Tsomokos4,Tsomokos5}, this paper considers electrons which are always in regions with zero electric and magnetic field.

\begin{figure}[htbp] 
   \centering
   \includegraphics[width=3in]{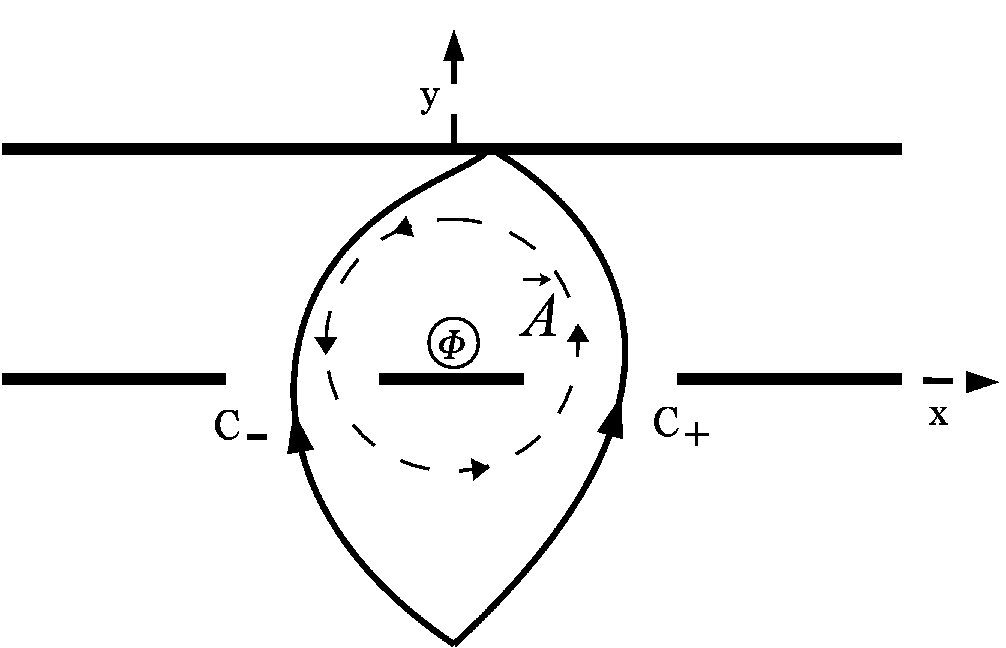} 
   \caption{Standard setup for Aharonov-Bohm effect with $x$ and $y$ directions indicated. The $z$-axis is ``up" and out of
   the plane of the page}
   \label{fig:ABFigure1}
\end{figure}

As Aharonov and Bohm \cite{Aharonov1959} showed, information about the magnetic flux (modulo a constant) through a solenoid can be measured via its effect on the interference pattern of electrons in a two-slit experiment as illustrated in Figure \ref{fig:ABFigure1}. For completeness and to fix notation, we briefly review the calculation in simplified form.

A particle of charge $q$ traversing the path $C_{\pm}$ in the presence of a magnetic vector potential $\vec{A}$ can be described by a wavefunction
\be
\psi_{\pm}(x)e^{\frac{iq}{\hbar c}\int_{C_{\pm}}\vec{A}\cdot d\vec{l}},
\ee

\noindent where $\psi_\pm(x)$ is the wavefunction in the absence of a vector potential, and the integral is taken along the corresponding path $C_\pm$.

We define \emph{flux parameter} $\phi$ as

\begin{equation}
\phi=\frac{q}{\hbar c}\left(\int_{C_+}\vec{A}\cdot d\vec{l}-\int_{C_-}\vec{A}\cdot d\vec{l}\right)=\frac{q}{\hbar c}\oint_{C} \vec{A}\cdot d\vec{l}=\frac{q}{\hbar c}\Phi,
\label{eq:fluxparameter}
\end{equation}
where $C$ is a closed curve created by following $C_{+}$ and returning along $C_{-}$, and $\Phi$ is the flux through the solenoid and thus through any closed surface bounded by $C$.

In terms of these variables, the total wavefunction at the screen is 
\be
\Psi(x,\phi)=e^{\frac{iq}{\hbar c}\int_{C_{+}}\vec{A}\cdot d\vec{l}}\left(\psi_{+}(x)+\psi_{-}(x) e^{-i\phi}\right),
\ee

\noindent which is gauge invariant up to an overall phase. This phase is unobservable in the interference pattern, which is proportional to $|\Psi(x,\phi)|^2$.

\section{Flux superposition}
In the standard A-B effect,  magnetic field inside the solenoid in Fig.\ref{fig:ABFigure1} can point in either the $+z$ or the $-z$ direction, resulting in a positive and negative flux, respectively. Suppose now that the current in the solenoid is in a superposition of two macroscopic states corresponding to equal and opposite currents, resulting in a superposition of positive and negative magnetic flux inside the solenoid. The case of superposition of fluxes of different magnitudes can also be studied, but the two physical systems being superposed then differ in their energies due to different quantities of energy in the corresponding magnetic fields. This would complicate matters since each state would evolve differently in time. Here, we limit ourselves to superpositions of two basis states with fluxes of equal magnitude but opposite sign.

Since the two basis states correspond to distinct eigenvalues of the Hermitian operator which determines the flux, they are orthogonal. We can then construct a wavefunction of an electron interacting with the superposition of the corresponding vector potentials.

The  probability amplitude for a particle to be found at position $x$ on the screen for flux  ``up'' due to vector potential $\vec{A}_\uparrow$ is
\begin{equation}
\label{ref:psiu}
\pu(x,\phi)=e^{\frac{iq}{\hbar c}\int_{C_{+}}\vec{A}_\uparrow\cdot d\vec{l}}\left(\psi_{+}(x)+\psi_{-}(x) e^{i|\p|}\right),
\end{equation}

\noindent while for the same flux ``down'' due to a vector potential $\vec{A}_\downarrow$ we have

\begin{equation}
\label{ref:psid}
\pd(x,\phi)=e^{\frac{iq}{\hbar c}\int_{C_{+}}\vec{A}_\downarrow\cdot d\vec{l}}\left(\psi_{+}(x)+\psi_{-}(x) e^{-i|\p|}\right).
\end{equation}

The ket describing the wavefunction of the particle in case of the superposed flux (amplitude to be ``up'' is $\cos(\theta/2)$ and to be ``down'' is $\sin(\theta/2)$) is
\be
\label{eq:generalsuppsi}
{\ket{\Psi(x,\p)}}=\cos{\frac{\A}{2}} \ket{\pu} +\sin{\frac{\A}{2}}e^{i\omega}\ket{\pd},
\ee
where $0\leq \A \leq \pi$  and $ 0\leq \omega \leq 2\pi$.
The probability distribution at the screen is
\be
\braket{\Psi(x,\p)}{\Psi(x,\p)}=\cos^{2}\frac{\A}{2} \braket{\pu}{\pu}+\sin^{2}\frac{\A}{2} \braket{\pd}{\pd}+\sin\frac{\A}{2}\cos\frac{\A}{2}\left[ e^{i\omega}\braket{\pu}{\pd}+e^{-i\omega}\braket{\pd}{\pu}\right]\label{ref:P}.
\ee

Since $\ket{\pu}$ and $\ket{\pd}$ are orthogonal (as noted above),
the cross terms in Eq.(\ref{ref:P}) vanish as is required in order to maintain gauge invariance (there must be no remaining dependence on $\vec{A}$). As a result, the interference pattern at the screen cannot tell us anything about $\omega$, the relative phase between the ``up'' and ``down'' states of the magnetic field. Remarkably, however, some information is still available. Dropping the explicit x-dependence, the total probability density at the screen is 

\be
|\Psi(x,\phi)|^{2}=|\psi_{+}|^{2}+|\psi_{-}|^{2}+2\R\left( \psi_{+}^{*} \psi_{-}\right) \cos{|\p|}
-2\Im\left(\psi_{+}^{*}\psi_{-}\right)\sin{|\p|}\left(\cos^{2}\frac{\A}{2}-\sin^{2}\frac{\A}{2}\right).
\label{eq:probability}
\ee

It is instructive to consider some limits of Eq.(\ref{eq:probability}). For $|\p|=0$, $|\Psi(x,\phi)|^{2}$  reduces  to the probability density for a particle passing through a double slit, as expected. The results for various choices of $\A$ are more interesting.
\begin{enumerate}
\item{For $\A=0$, $|\Psi(x,\phi)|^{2}$  goes to regular A-B effect with an ``up" flux.}
\item{For $\A=\pi$, $|\Psi(x,\phi)|^{2}$  goes to regular A-B effect with a ``down" flux.}
\item{For $\A=\pi/2$, the last term in Eq.(\ref{eq:probability}) goes to zero. There is still an interference pattern, but it is {\it{different}} from the interference patterns in cases 1 and 2 above.}
\end{enumerate}

The first two limits indicate that that Eq.(\ref{eq:probability}) reduces to the expected expressions for two classical flux states.  More interesting is the third limit, which indicates that it is in principle possible to extract information about the quantum mechanical state of magnetic flux in the A-B experiment from the electron diffraction pattern without disturbing the state of the flux. This information is extracted via a fundamentally nonlocal operation involving the vector potential over an extended region of spacetime and without any interaction in the region where the (superposition of) classical magnetic fields are present (i.e., the excluded region inside the solenoid).  In particular, one does not need to involve any local interaction Hamiltonian involving the magnetic field $\vec{B}$ of the form $\vec{\sigma} \cdot \vec{B}$ to describe how this process takes place. 

While information is obtainable from the interference pattern, it is not complete. In the classical A-B effect, the flux $\Phi$ can only be determined modulo $2\pi\hbar c/e$. For $\Phi=0$ modulo $2\pi\hbar c/e$, the interference pattern is the same as that for a positive and negative classical flux -- there is no effect on the pattern.

Superpositions of two magnetic fluxes which would individually be detectable via the A-B effect can give rise to interference patterns, which differ from any found in the classical (or nonsuperposed) case.

Figure \ref{fig:ABFigure3}(a)-(c) in the following section shows these patterns as a function of the magnitude of the flux parameter. They clearly indicate that the pattern produced by a superposition of equal magnitude fluxes in both directions is different from that produced in the usual A-B effect except for when the A-B effect would not detect either flux anyway (flux is zero modulo $2\pi\hbar c/e$).

Note that the information about superposition is gathered from the full interference pattern and is  built up, little by little, with contributions from each electron. For any finite experimental resolution and finite number of electrons scattered, one can only construct the likelihood that the observed interference pattern corresponds to the prediction for an arbitrary superposition of fluxes. No single electron scattering event provides unambiguous information even about the flux modulo $2\pi\hbar c/e$.

One can also consider the case of fluxes, which are detectable by the usual A-B effect, but now with unequal superpositions of ``up" and ``down" fluxes ($|\cos\theta/2|^2 \neq |\sin\theta/2|^2$). In this case, again, one extracts information about the nature of the state without any interaction, which should cause the ``collapse'' of the state into one of definite flux.

\section{Experimental Predictions}
\label{Experimental Predictions}
\begin{figure}[htbp] 
   \centering
   \includegraphics[width=3in]{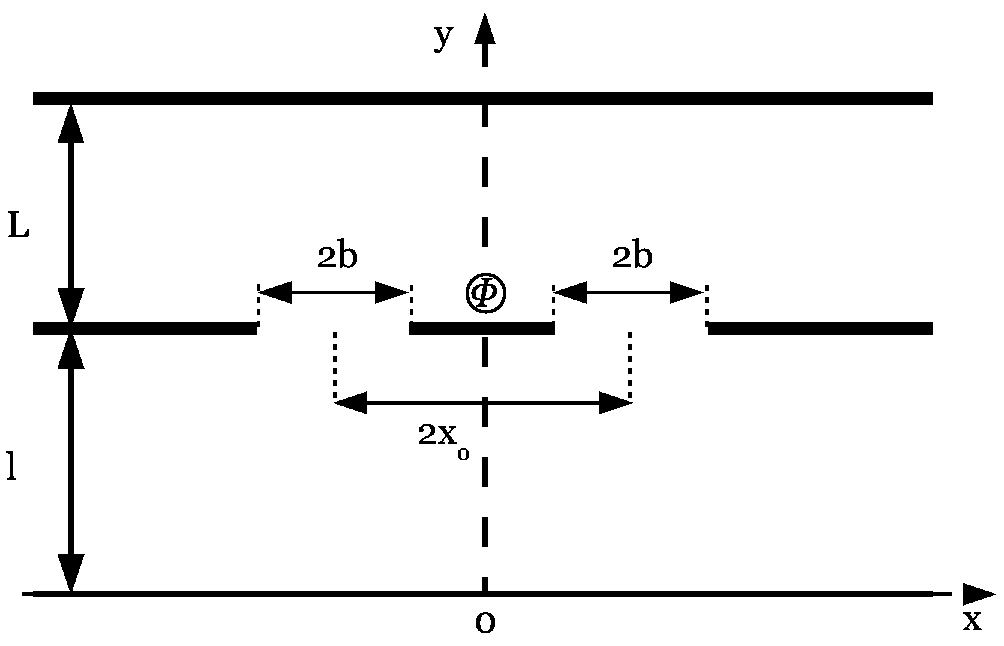} 
   \caption{A possible experimental setup.}
   \label{fig:ABFigure2}
\end{figure}

We next consider what might actually be observed in more realistic experimental setting, following the work of Aguilera-Navarro and Quick \cite{Navarro}, in which they use the Feynman 
path integral approach to calculate the A-B interference pattern for electrons passing through a 
barrier with two finite width slits (see Fig. \ref{fig:ABFigure2}). 

In their approximation, electron motion in the $y$-direction is classical. The wavefunction at the screen of an electron passing through the right and left slits is
\be
\psi_{\pm}(x)=-\frac{iN}{\sqrt{\mathcal{N}}}e^{i\frac{\pi x^{2}}{\lambda(L+l)}}\left[Ei\left(\beta\left(x_{0}+b\mp\frac{l}{L+l}x\right)\right)-Ei\left(\beta\left(x_{0}-b\mp\frac{l}{L+l}x\right)\right)\right].
\label{eq:NQpsi2}
\ee
Parameters $N$ and $\beta$ are defined as
\be
N=\sqrt{\frac{1}{2\lambda(l+L)}}
\ee
and
\be
\beta=\sqrt{\frac{2}{\lambda}\left(\frac{1}{l}+\frac{1}{L}\right)}.
\ee
The normalization factor $\mathcal{N}$ is 
\be
\mathcal{N}=\frac{4b}{\lambda l}.
\ee
Here, $\lambda=h/mv$ is the de Broglie wavelength of the electron, where $m$ is the mass of the electron, and $v$ is its speed in the $y$-direction which is constant. $Ei(z)$ represents the complex Fresnel integral defined as
\be
Ei(z)\equiv \int_{0}^{z}e^{i\frac{\pi}{2}\eta^{2}}d\eta.
\ee

Still following Aguilera-Navarro {\em et al.} ({\em op. cit.}) we use the same parameter values as were used by J\"{o}nsson \cite{Jonsson}. The distance between the source and the slit is taken to be $l = 10\mbox{m}$, the slit to screen distance $L = 1\mbox{m}$, electron wavelength $\lambda = 5\times10^{-12}\mbox{m}$, the slit width parameter is $b =0.25\times10^{-6}\mbox{m}$,  and the slit separation parameter is $x_{0}= 10^{-6}\mbox{m}$ (See Fig.{\ref{fig:ABFigure2}). 

Using these values in the expressions for $\psi_{\pm}(x)$ in Eq.(\ref{eq:probability}), we generate three plots of the probability density at the screen as a function of $x$ and $\p$. These are shown in Fig.\ref{fig:ABFigure3}. Figures \ref{fig:ABFigure3}(a) and \ref{fig:ABFigure3}(b) show the results of the standard A-B effect for various magnitudes of the flux parameter. Figure  \ref{fig:ABFigure3}(a) corresponds to a positive flux, or $\theta=0$. It shows a double slit pattern shifting to the left as a function of increasing flux magnitude, as is expected, because the electron has negative charge. Figure  \ref{fig:ABFigure3}(b) corresponds to a negative flux, or $\theta=\pi$ and it shows an expected shift to the right. Finally, Fig. \ref{fig:ABFigure3}(c), corresponding to a superposed state of the flux, or $\theta=\pi/2$, is distinctly different from the first two, most notably when the value of $\phi$ is $\pi/2$. Note that there is no net shift of the center-of-mass of the distribution as a function of flux, as shown in reference \cite{Navarro}, which clarified earlier concerns about whether or not there was a net average transfer of momentum between the magnetic field source and the electrons.
\begin{figure}[htbp] 
 \includegraphics[height=1.7in]{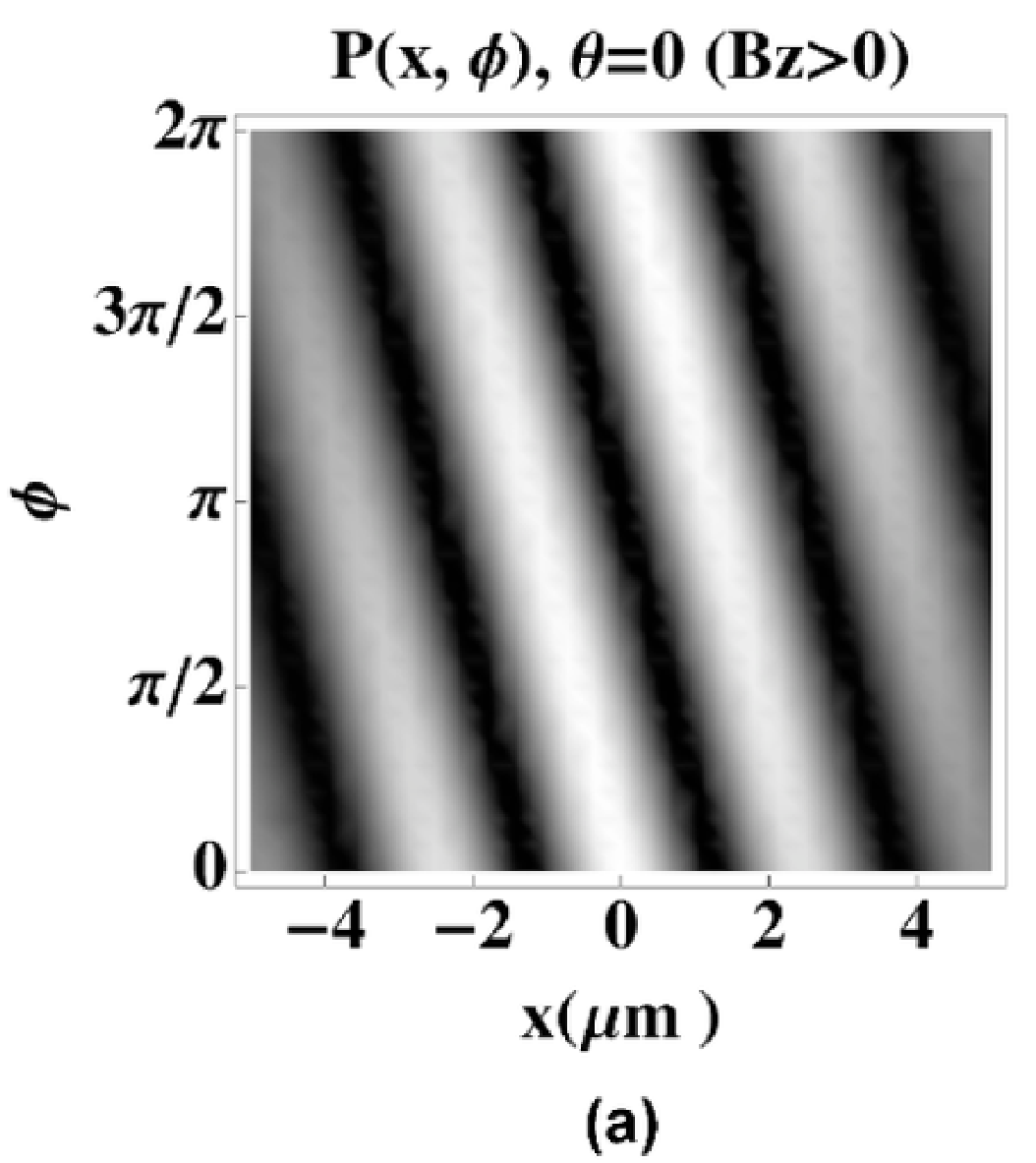}\includegraphics[height=1.7in]{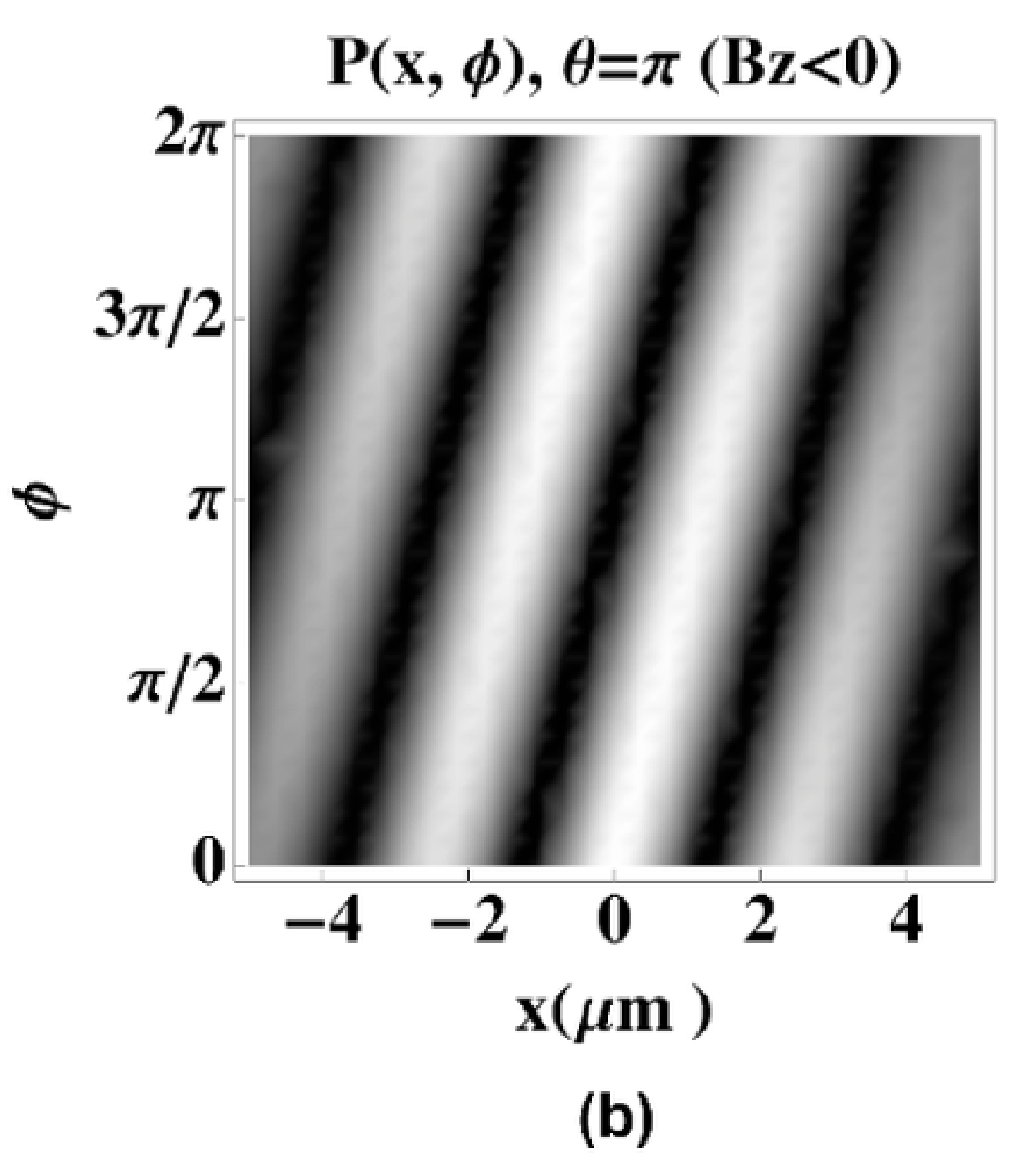}\includegraphics[height=1.7in]{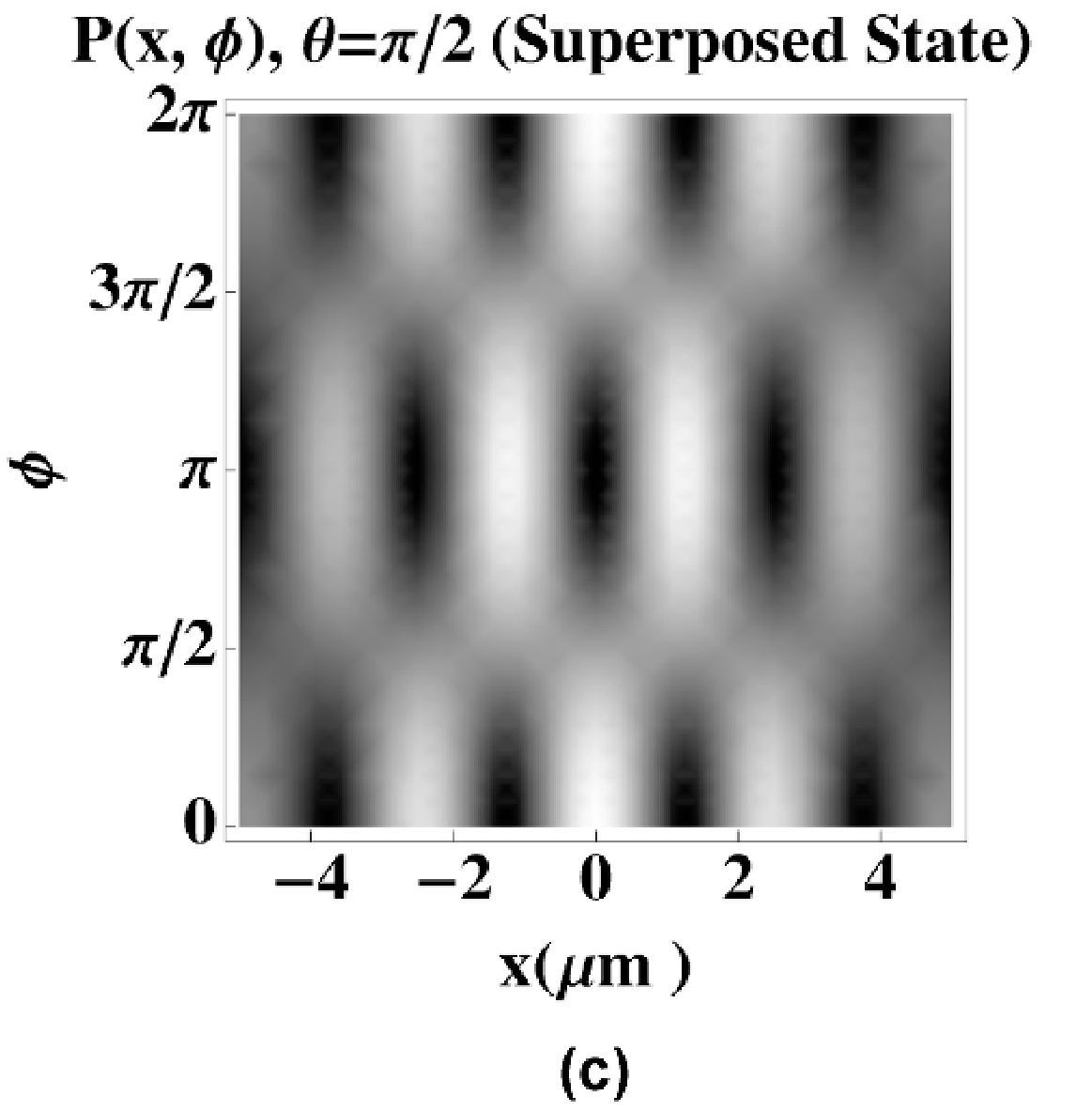}
   \caption{Density plot of the probability density at the screen for $0\leq\phi\leq2\pi$ for $\theta=0, \pi$, and $\pi/2$. Note that one can clearly distinguish various values of $\phi$ for $\theta=\pi/2$ and that this information is obtained without disturbing the state of the flux}
   \label{fig:ABFigure3}
\end{figure}

Figure \ref{fig:ABFigure4} shows how different $\theta$ values shift a pattern produced by a positive flux into one produced by a negative flux.
\begin{figure}[htbp] 
 \includegraphics[height=1.7in]{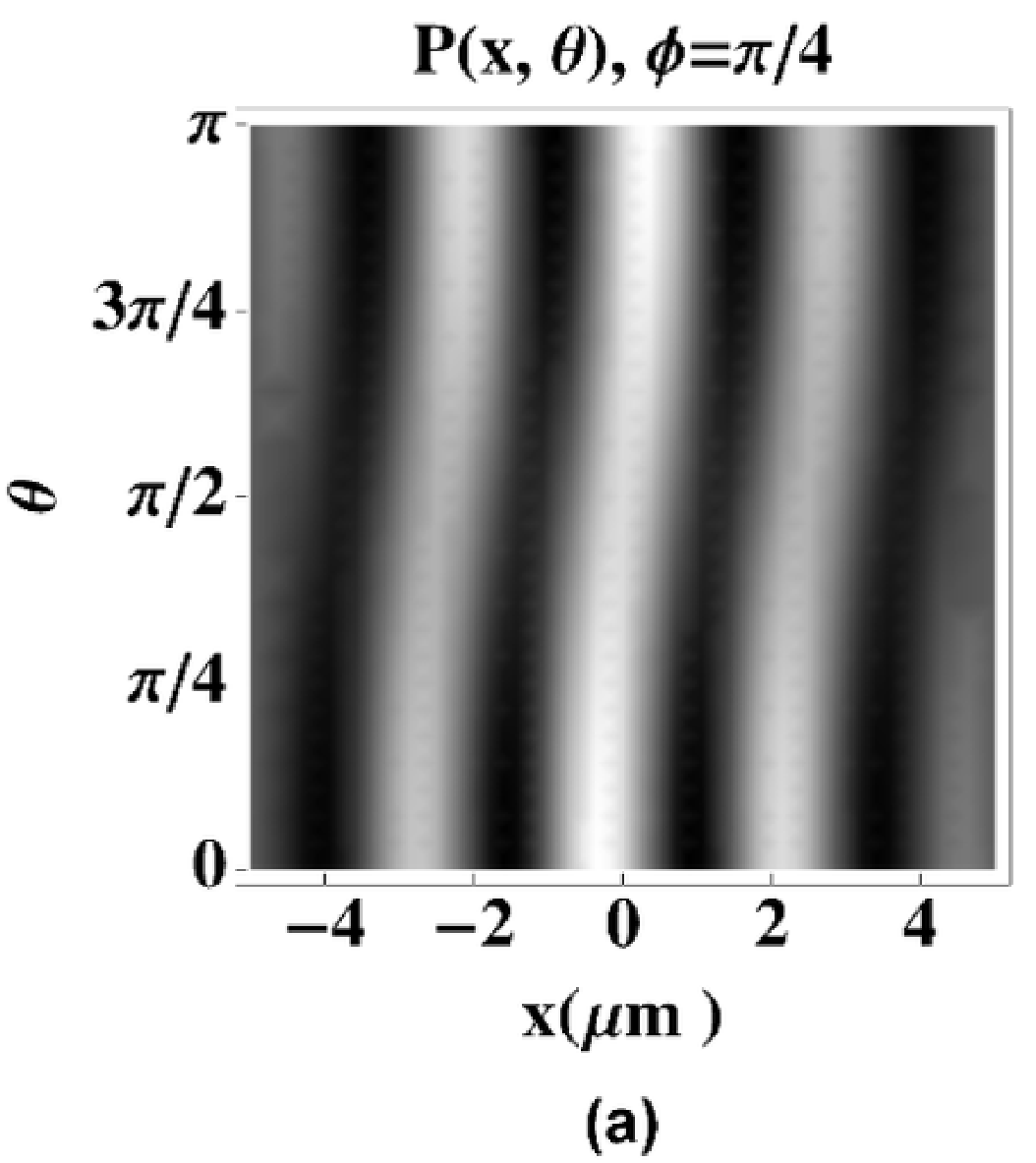}\includegraphics[height=1.7in]{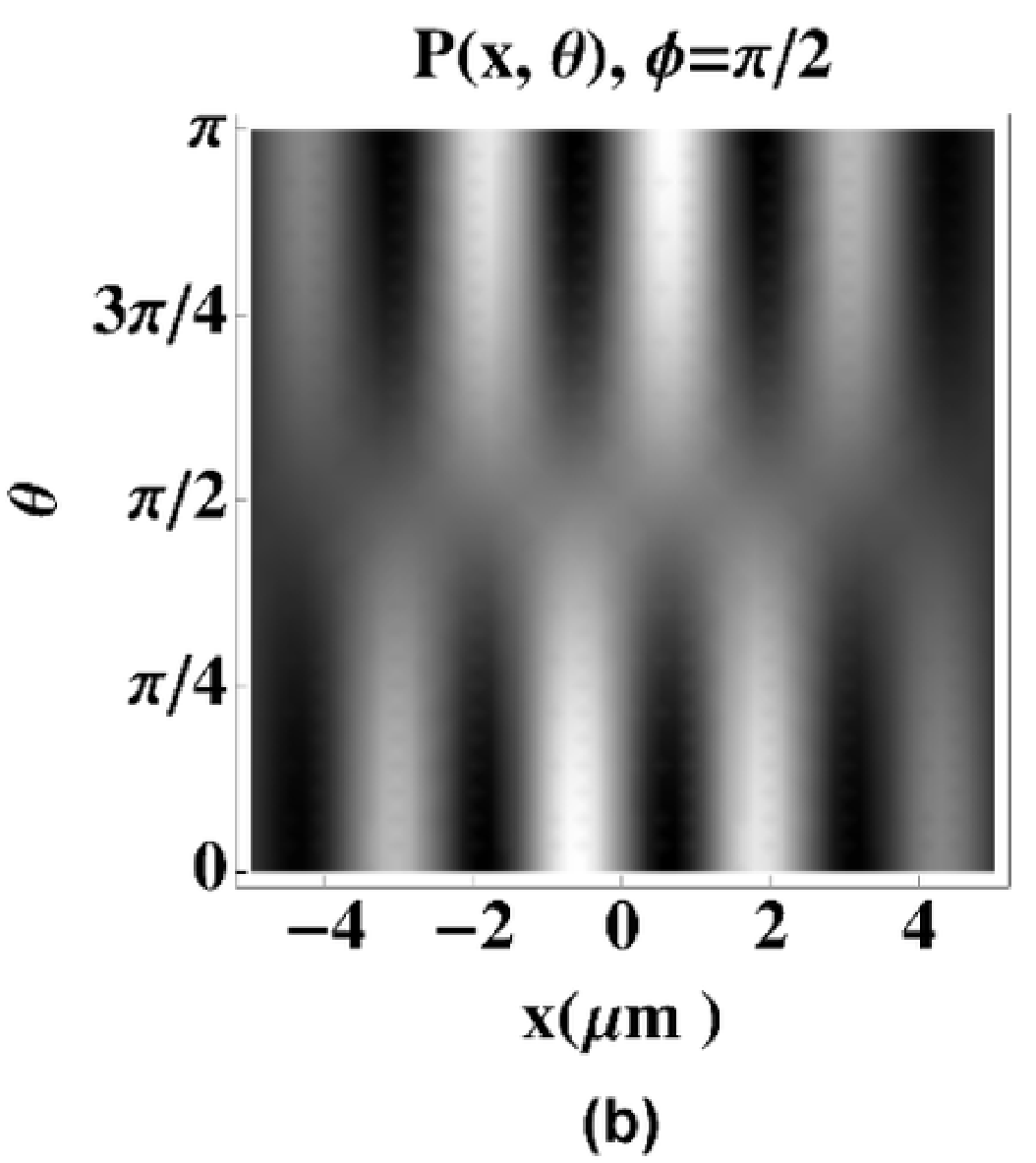}\includegraphics[height=1.7in]{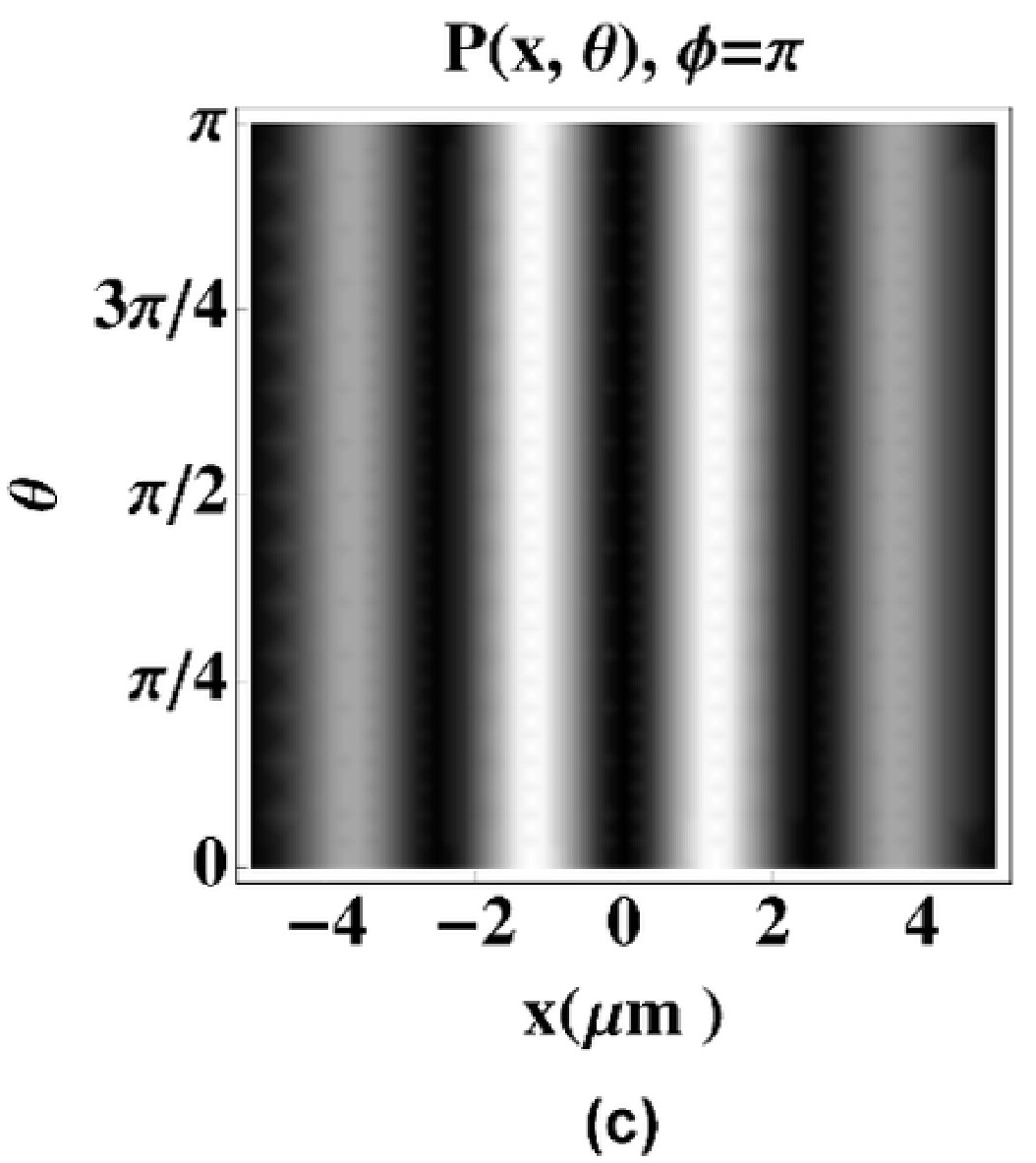}
   \caption{Density plot of the probability density at the screen for $0\leq\theta\leq\pi$ for $\phi=\pi/4, \pi/2$, and $\pi$}
   \label{fig:ABFigure4}
\end{figure}

The question of how one could obtain a superposed flux for such an experiment is somewhat more involved, but there is cause for optimism. van der Wal {\em et al.} \cite{vanderWal2000} and Friedman {\em et al.} \cite{Friedman} have demonstrated quantum superposition of macroscopic persistent-current states in superconducting rings, which would produce superpositions of magnetic fluxes through the holes in the rings. 

Symmetric and antisymmetric superpositions of classical states with persistent current of 0.5 mA corresponding to the center-of-mass motion of millions of Cooper pairs have been achieved, so it is certainly not inconceivable that an experimental test of the predictions made here could be achieved. We hope to return to this issue in a future paper.

\section{Conclusion}
We have shown that for a generalization of the A-B experiment involving a superposition of equal amounts of flux of opposite signs, the interference pattern can determine that the quantum system is in a superposition without disturbing it. The nature of the process by which this information is extracted is interesting since it is by nature highly nonlocal, and connects the well-known non-locality in quantum mechanics ({\em i.e.} EPR effect \cite{EPR}) with non-nonlocality of the vector potential description of electromagnetism.

In most quantum mechanical measurement situations, one discovers that a state is in a superposition of two basis states from the accumulated statistics from measurements on identically prepared states.  For example, a photon polarized along the $x$-axis can be described as a superposition of two states along the $z$ and $-z$ axes, but a single measurement in the $z$ direction will give one of two values, either of which could have been arisen from a photon initially polarized along the corresponding direction in $z$. Once the measurement is done, the initial state has to be prepared again in order to learn anything about whether or not it was in a superposition. Finding that a repetition of the experiment yields a different polarization along the $z$ axis, one is really learning not so much about the state of a given photon, but about a given technique to prepare photons in that state.

Here we show that it is possible to prepare a state which can be experimentally identified as being in a superposition of basis states without collapsing it into one or the other. That information is obtained from repeated electron scatterings. Each provides a small amount of information, but none of them affects the state being studied and the state need not be prepared more than once.

The calculations presented here are done in the framework of standard quantum mechanics. It is interesting to note that in hidden variable theories where the flux is actually has one of the two values in the superposition, but one just doesn't know which, the interference pattern would be that for the corresponding classical flux. As such, the experiments proposed here test hidden variable theories \cite{Genovese2005} without the need to invoke Bell inequalities. The only statistical feature involved is that the degree of certainty about which interference pattern is actually seen depends on the number of scattered electrons contributing to it.

Experimental possibilities were studied, and while the technical details are challenging, it does not seem beyond the realm of possibility that these ideas could be tested in the laboratory. Indeed, this effect could find applications in quantum computing \cite{Aristov2008} for determining information about the states of qubits \cite{vanderWal2000,Wilhelm2003,Orlando2002194,Orlando2002294,Tian2000,Orlando1999,Mooij1999} stored as superpositions of magnetic fluxes. We thank an anonymous referee for suggesting that the electron beam interference part of this experiment could be replaced with electrons passing through a Josephson junction in a superconducting quantum interference device (SQUID) enclosing a quantum flux.

While our calculations indicate that these ``A-B-interactions'' would not force the collapse of the wavefunction describing the flux into an eigenstate of flux, what actually happens in practice is an experimental question. It should be emphasized that the information about the state of the flux comes from the interference pattern and that very little information can be obtained from observing the pattern with only one or a few electrons.

It would be interesting if these measurements did indeed change the state of the flux, either all-at-once at some point, say after some number $N$ of electron scatterings, or gradually over time with an initial superposition smoothly evolving to a nonsuperposed state without any ``jumps." If the state of the flux were entangled with any other quantum system, its state could also be tracked in a similar fashion.

For  experiments where the pattern is built up slowly and the information on the flux is observed building up over time, the location in which each electron is detected gives a contribution to the likelihood \cite{Edwards1992} that the distribution obtained thus far comes from the theoretically calculated distribution. The likelihood distribution for $\theta$ could, in principle, be followed over time for signs of a jump between states or a smooth evolution from one to another, if indeed the state changes due to these A-B measurements.

This system can also be considered as a toy model for the quantum mechanical propagation of a particle in a background spacetime which is a superposition of different classical geometries. Here one has a situation equivalent to a superposition of two spacetimes with oppositely spinning cosmic strings \cite{Mazur1986,Samuel1987} which are flat outside the string.

As a more physical  example, one could consider the scattering of particles around spinning objects that drag spacetime around them (the Lense-Thirring effect). If such objects are in a superposition of two spin states, the background spacetime would have to be treated as a superposition of two classical geometries. In this case, however, and in contrast with the A-B effect or the spinning string considered above, the particles would move in regions where they would feel classical forces. The situation described here is probably the simplest similar system that can be studied in detail.

\begin{acknowledgments}
K. B. would like to thank John Stachel for his continuous support, encouragement and many useful discussions.  J. S. is supported in part by NSF grant PHY-0855388. Both authors would like to thank the anonymous referee for valuable comments on an earlier version of this paper. 
\end{acknowledgments}

\bibliography{ABOctober2013.bib}
\end{document}